\documentclass[10pt,onecolumn,aps,prd,preprintnumbers,showpacs,superscriptaddress,nofootinbib,amsmath,amssymb,floats,floatfix,showkeys,notitlepage,longbibliography]{revtex4-1}
\usepackage{graphicx}	
\usepackage{amssymb}
\usepackage{dcolumn}
\usepackage{mathtools}
\usepackage{amsmath}
\usepackage{xcolor}
\usepackage{color}
\usepackage{theorem}
\usepackage{subfigure, rotating, bm, array}
\usepackage[pagebackref=false, colorlinks=true]{hyperref}
\hypersetup{linkcolor=blue, citecolor=blue,urlcolor=blue} 

\usepackage[utf8]{inputenc}   
\usepackage[T2A]{fontenc}
\usepackage[english,russian]{babel}           

\begin{document}

\title{Second-Order Effects in Gravitational Wave Spacetime}

\author{Misyura M. A.}
\affiliation{Department of High Energy and Elementary Particles Physics, Saint Petersburg State University, University Emb. 7/9, Saint Petersburg, 199034, Russia \\
IPGG RAS, Makarova Emb. 2, Saint Petersburg, 199034, Russia}
\email{max.misyura94@gmail.com}

\begin{abstract}

In this paper, we investigate gravitational waves beyond the linear approximation, focusing on second-order contributions sourced by linearized waves in the transverse-traceless (TT) gauge. A general spacetime metric is constructed, and both timelike and null geodesic congruences are analyzed. For the timelike congruence, a non-vanishing expansion scalar and shear tensor are obtained, while the rotation tensor is found to vanish. In contrast, all these quantities vanish for the null congruence. Using a parallel-transported orthonormal tetrad, we derive the geodesic deviation equations up to second order in the wave amplitude $H$, showing that, as in the linear case, stretching and compression occur in the transverse $x$ and  $y$  directions. However, when solving the geodesic equations numerically within the 3+1 formalism, we observe an additional effect: test particles undergo a small, second-order displacement along the direction of wave propagation.

\end{abstract}
\maketitle

\section{Introduction} 

Gravitational waves are ripples in the fabric of spacetime, first predicted theoretically over a century ago. Their direct detection, however, occurred only recently, in 2015, at the LIGO gravitational wave observatories~\cite{bib:gw1}. As of today, three such observatories are in operation, and the resulting catalog contains nearly one hundred events, including mergers of black holes and neutron stars~\cite{bib:gw2,bib:gw3,bib:gw4,bib:gw5}.

In the near future, a new space-based observatory is planned, with the aim of detecting gravitational waves emitted by supermassive black holes, extreme mass ratio inspirals (EMRIs), and potentially cosmological sources such as those originating from the early Universe and the Big Bang epoch~\cite{bib:lisa}. This development opens a new observational window into the properties, population, and distribution of astrophysical sources of gravitational radiation.

The detection of gravitational waves relies on their interaction with matter~\cite{bib:intgw}, and different observational techniques are used depending on the frequency range of the wave~\cite{bib:romangw,bib:chigw}. However, all existing methods are grounded in the linearized theory of gravitational waves and are sensitive to effects that scale with the first order of the wave amplitude. Among these is the geodesic deviation effect~\cite{bib:meggori}, which has been thoroughly investigated within linearized theory in various scenarios~\cite{bib:grishuk,bib:dev1,bib:dev2}.

It is well known that general relativity is a fundamentally nonlinear theory. This naturally raises the question: is it possible to observe nonlinear manifestations of gravitational waves? In other words, how does the interaction between gravitational waves and matter change if the wave amplitude exceeds the regime currently accessible to modern interferometric detectors?

The investigation of nonlinear gravitational waves began shortly after the development of the linearized theory. Initially, Einstein himself questioned their existence due to apparent singularities in early solutions. However, detailed work in the 1960s by Trautman, Bondi, Pirani, Robinson, Sachs, and others~\cite{bib:gwingr,bib:gwingr2,bib:gwingr3,bib:gwingr4,bib:gwingrex1,bib:gwingrex2}—including the study of exact wave-like solutions and the asymptotic structure of spacetime at null infinity—established gravitational waves as a physical reality and led to the eventual development of experimental detection efforts.
Despite this progress, a clear and universally accepted definition of what constitutes a strong gravitational wave remains elusive. When the amplitude of a wave approaches unity, the distinction between the wave and the background metric becomes ambiguous; the spacetime itself evolves in a fully nonlinear manner. As a result, most studies of nonlinear gravitational waves and their effects rely on perturbative expansions around various background metrics~\cite{bib:non1,bib:non2,bib:non3,bib:non4}.

The aim of the present work is to study nonlinear effects that emerge at second order in the amplitude of linearized gravitational waves. These quadratic contributions can be associated with the wave's effective energy, but the analysis is confined to a region within a single wavelength.

This article is devoted to the study of spacetime distortions produced by second-order gravitational waves, where the source is taken to be waves from linearized theory. The structure of the paper is as follows:
In Section~2, we briefly review the linearized theory of gravitational waves. In Section~3, using the formalism of relativistic perturbation theory, we derive the second-order gravitational wave equations and introduce the corresponding metric tensor. In Section~4, we study a congruence of geodesics and analyze the geodesic deviation equation. We also numerically solve the geodesic equations within the \( 3+1 \) formalism. Section~5 presents our conclusions.
Throughout the paper, we adopt the metric signature \( (-,+,+,+) \), and use natural units where \( c = 1 \) and \( G = 1 \), unless otherwise specified.

\section{The linearized theory}

The linearized theory of gravity is based on expanding the Einstein field equations around a flat Minkowski background:
\begin{equation}
g_{\alpha \beta}=\eta_{\alpha \beta}+h_{\alpha \beta}
\end{equation}
where \( h_{\alpha \beta} \) is treated as a small perturbation, and indices are raised and lowered using the background metric \(  h = \eta^{\alpha \beta} h_{\alpha \beta} \). Also it is convenient to define the trace-reversed perturbation:
\begin{equation}
h_{\alpha \beta} = \bar{h}_{\alpha \beta} - \frac{1}{2} \eta_{\alpha \beta} \bar{h},
\end{equation}
where \( \bar{h} = \eta^{\mu\nu} \bar{h}_{\mu\nu} \). The general relativity is invariant under infinitesimal coordinate transformations of the form \( x^{\mu} \rightarrow x'^{\mu} = x^{\mu} + \xi^{\mu}(x) \), where \( \xi^{\mu} \) is an arbitrary vector field. Under such a transformation, the metric perturbation transforms as
\begin{equation}
    h'_{\alpha \beta}= h_{\alpha \beta}+\mathcal{L}_{\xi}
    \eta_{\alpha \beta} = \partial_{\beta} \xi_{\alpha}+\partial_{\alpha} \xi_{\beta}+h_{\alpha \beta} \label{eq:diff1}
\end{equation}
where \( \mathcal{L}_\xi \) denotes the Lie derivative with respect to \( \xi^{\mu} \). In terms of the trace-reversed perturbation, the transformation becomes
\begin{equation}
  \bar{h}'_{\alpha \beta} = \partial_{\beta} \xi_{\alpha}+\partial_{\alpha} \xi_{\beta}+\bar{h}_{\alpha \beta} - \eta_{\alpha \beta} \,
      \partial_{\mu}\! \left(\xi^{\mu} \right)
\end{equation}
This gauge freedom allows us to impose the Lorentz (harmonic) gauge condition:
\begin{equation} \label{eq:lor1}
\partial^{\beta} \bar{h}_{\alpha \beta} = 0.
\end{equation}
Assuming a vacuum region where the energy-momentum tensor vanishes, \( T_{\alpha \beta} = 0 \), the linearized Einstein equations reduce to the wave equation:
\begin{equation}
\square \, \bar{h}_{\alpha \beta} = 0.
\end{equation}
Even within the Lorentz gauge, residual gauge transformations are allowed provided the vector field \( \xi^\mu \) satisfies the homogeneous wave equation \( \square \xi^\mu = 0 \). Under such transformations, the induced change in \( \bar{h}_{\alpha \beta} \) is
\begin{equation} \label{eq:resg1}
\xi_{\alpha \beta} = \partial_{\alpha} \xi_{\beta} + \partial_{\beta} \xi_{\alpha} - \eta_{\alpha \beta} \partial_{\mu} \xi^{\mu},
\end{equation}
which also satisfies \( \square \, \xi_{\alpha \beta} = 0 \).

Thus, by imposing the Lorentz gauge condition~\eqref{eq:lor1} and applying the residual gauge transformation~\eqref{eq:resg1}, one arrives at the transverse-traceless (TT) gauge, in which the gravitational wave possesses two independent polarization components. For example, if the wave propagates along the \( z \)-axis with frequency \( \omega \), the corresponding perturbation to the metric takes the form:
\begin{equation}
g_{\alpha \beta}=\left(\begin{array}{cccc}
-1 & 0 & 0 & 0
\\
 0 & 1+\mathit{H} \cos\! \left(\omega \left(t-z\right)\right) & \mathit{H} \cos\! \left(\omega \left(t-z\right)\right) & 0
\\
 0 & \mathit{H} \cos\! \left(\omega \left(t-z\right)\right) & 1-\mathit{H} \cos\! \left(\omega \left(t-z\right)\right) & 0
\\
 0 & 0 & 0 & 1
\end{array}\right)\label{eq:cosh}
\end{equation}
where \( H \) denotes the wave amplitude.

It is important to note that in such a spacetime, a test particle initially at rest remains at rest in the coordinate system even after the wave has passed. However, when viewed in a proper detector frame and taking into account the invariance of the linearized Riemann tensor under infinitesimal coordinate transformations of the form~\eqref{eq:diff1}, one recovers the well-known effect of geodesic deviation.

\section{Second order gravitational waves} 

To investigate gravitational waves beyond the linear approximation, we consider a perturbative expansion of the metric tensor up to second order in the wave amplitude \( H \):
\begin{equation} 
g_{\alpha \beta} = \eta_{\alpha \beta} + H h_{\alpha \beta} +
\frac{H^2}{2} h^{(2)}_{\alpha \beta}, \label{eq:mett2}
\end{equation}
where \( h^{(2)}_{\alpha \beta} \) denotes the second-order perturbation.
As in the linear case, we define the trace-reversed second-order perturbation:
\begin{equation}
h^{(2)}_{\alpha \beta} = \bar{h}^{(2)}_{\alpha \beta} - \frac{1}{2} \eta_{\alpha \beta} \bar{h}^{(2)},
\end{equation}
where \( \bar{h}^{(2)} = \eta^{\mu \nu} \bar{h}^{(2)}_{\mu \nu} \). To impose the Lorentz gauge condition at second order,
\begin{equation}
\partial^{\alpha} \bar{h}^{(2)}_{\alpha \beta} = 0,
\end{equation}
we must understand how \( h^{(2)}_{\alpha \beta} \) transforms under arbitrary coordinate transformations of the form \( x^{\mu} \rightarrow x'^{\mu}(x) \), i.e., under general gauge transformations.

This analysis can be formalized by performing a Taylor expansion of tensor fields on a manifold. The infinitesimal coordinate transformation up to second order in a small parameter \( \lambda \) takes the form
\begin{equation} \label{eq:ipt2}
x'^{\mu} = x^{\mu} + \lambda \xi^{\mu}_{(1)} + \frac{\lambda^2}{2}
\left( \xi^{\mu}_{(1),\nu} \xi^{\nu}_{(1)} + \xi^{\mu}_{(2)} \right),
\end{equation}
where \( \xi^{\mu}_{(1)} \) and \( \xi^{\mu}_{(2)} \) are independent vector fields generating a one-parameter family of diffeomorphisms \( \Phi_\lambda: \mathcal{M} \rightarrow \mathcal{M} \), such that \( \Phi_\lambda(p) = q \), with \( \Phi_0(p) = p \), and \( p, q \in \mathcal{M} \) lie within a common coordinate neighborhood. The parameter \( \lambda \) may be introduced naturally from the physical context, for example, as an expansion parameter related to the amplitude of perturbations. A more detailed information can be found in~\cite{bib:g1,bib:g2,bib:g3,bib:g4} and references therein.
From equation~\eqref{eq:ipt2}, the transformation of a generic tensor field \( T \) under a second-order gauge transformation is given by
\begin{equation}
T'(\lambda) = T(\lambda) + \lambda \mathcal{L}_{\xi_{(1)}} T +
\frac{\lambda^2}{2} \left( \mathcal{L}^2_{\xi_{(1)}} +
\mathcal{L}_{\xi_{(2)}} \right) T\,.
\end{equation}
 Expanding both \( T(\lambda) \) and \( T'(\lambda) \) perturbatively as
\begin{equation}
\begin{split}
T(\lambda) &= T_0 + \lambda\, \delta T + \frac{\lambda^2}{2} \delta^2 T\,, \\
T'(\lambda) &= T_0 + \lambda\, \delta T' + \frac{\lambda^2}{2} \delta^2 T'\,,
\end{split}
\end{equation}
we obtain the gauge transformations for the first- and second-order perturbations as
\begin{equation}
\begin{split}
\delta T' &= \delta T + \mathcal{L}_{\xi_{(1)}} T_0\,, \\
\delta^2 T' &= \delta^2 T + 2 \mathcal{L}_{\xi_{(1)}} \delta T +
\mathcal{L}^2_{\xi_{(1)}} T_0 + \mathcal{L}_{\xi_{(2)}} T_0\,.
\end{split}
\end{equation}
The first expression is the familiar result for linear perturbations, while the second captures the general structure of gauge transformations at second order.
Let us simplify the notation by replacing \( \xi^{\mu}_{(1)} \) with \( \xi^{\mu} \), which generates first-order gauge transformations. Then the second-order transformation becomes:
\begin{equation}
h'^{(2)}_{\alpha \beta}= h^{(2)}_{\alpha \beta} + 2 \mathcal{L}_{\xi}
h_{\alpha \beta} +
\mathcal{L}^2_{\xi} \eta_{\alpha \beta} +  \mathcal{L}_{\xi_{(2)}} \eta_{\alpha \beta} \,.
\end{equation}
After expanding all Lie derivatives, the previous expression reads as
follows
\begin{multline}
h'^{(2)}_{\alpha \beta}= h^{(2)}_{\alpha \beta}+ \partial_{\alpha} \xi_{\mu} \left( 2 \partial_{\beta} \xi^{\mu} +\partial^{\mu} \xi_{\beta}+2
    h^{\mu}_{\beta} \right) +  \left( 
    \partial_{\mu}h_{\alpha \beta}+  \partial_{\alpha}\partial_{\mu} \xi_{\beta} \right) 2
\xi^{\mu}
+\partial_{\beta}\xi^{\mu}
\partial_{\mu}\xi_{\alpha} + \\ +2 \partial_{\beta}\xi_{\mu}
 h_{\alpha}^{\mu}
+\partial_{\beta}\xi^{(2)}_{\alpha}
+\partial_{\alpha}\xi^{(2)}_{\beta} \,.
\end{multline}
Now, taking into account that \( h_{\alpha \beta} \) satisfies the Lorentz gauge condition and is traceless, i.e., \( \partial^{\beta} \bar{h}_{\alpha \beta} = 0 \) and \( \bar{h}_{\alpha \beta} = h_{\alpha \beta} \), we can analyze how the second-order quantity \( \partial^{\beta} \bar{h}^{(2)}_{\alpha \beta} \) transforms under gauge transformations. The result reads:
\begin{multline}
  \partial^{\beta} \bar{h'}^{(2)}_{\alpha \beta} = 2 \partial_{\beta}\xi^{\lambda}
    \partial_{\lambda}\partial^{\beta}\xi_{\alpha}
    +\xi^{\beta}
        \partial_{\beta}\square 
    \xi_{\alpha}  + \left(-2
        \partial_{\alpha}
    h_{\lambda}^{\beta}+2
        \partial^{\beta}h_{\alpha \lambda}\right)
    \partial_{\beta} \xi^{\lambda} + \left(2
        h_{\alpha \beta}+2 \partial_{\alpha} \xi_{\beta}+\partial_{\beta}\xi_{\alpha}\right) \square \xi^{\beta} +\\ + 2
        \partial_{\beta} h_{\alpha \lambda} 
    \partial^{\lambda} \xi^{\beta}  + \square \xi^{(2)}_{\alpha} +\partial^{\beta}\bar{h}_{2 \alpha \beta} \,.
\end{multline}
Thus, one can choose four arbitrary functions \( \xi^{(2)}_{\alpha} \) such that the second-order perturbations also satisfy the Lorentz gauge condition.
Next, we substitute the metric expansion~\eqref{eq:mett2} into the Einstein field equations. Neglecting all terms proportional to \( h \cdot h^{(2)} \), which are of higher order \( \mathcal{O}(H^3) \), and using the fact that the first-order perturbation satisfies the vacuum wave equation, \( \square h_{\alpha \beta} = 0 \), we obtain the following equations:
\begin{multline}
     \square \bar{h}^{(2)}_{\mu \nu}= 2 h^{\alpha\beta} \left( \partial_\alpha \partial_\beta h_{\mu\nu} + \partial_\mu \partial_\nu h_{\alpha\beta} - \partial_\alpha \partial_\mu h_{\beta\nu} - \partial_\alpha \partial_\nu h_{\beta\mu} \right) \\ 
+ 2 \left( \partial_\alpha h_\nu^\beta \, \partial^\alpha h_{\beta\mu} - \partial^\alpha h_{\nu\beta} \, \partial^\beta h_{\alpha\mu} \right) 
+ \partial_\mu h_{\alpha\beta} \, \partial_\nu h^{\alpha\beta} 
- \frac{1}{2} g_{\mu\nu} \partial_\alpha h^{\beta\kappa} \, \partial^\alpha h_{\beta\kappa} \,.
\end{multline}
Substituting the explicit form of the first-order gravitational wave \( h_{\alpha \beta} \) given in equation~\eqref{eq:cosh} into the second-order field equations, we obtain the source terms for the components of \( \square \bar{h}^{(2)}_{\alpha \beta} \)
\begin{equation}
  \square \bar{h}^{(2)}_{\alpha \beta} =  \left(\begin{array}{cccc}
-2   w^{2} \left(1+3 \cos\! \left(2 \omega \left(t-z\right)\right)
\right) & 0 & 0 & 2   w^{2} \left(1+3 \cos\! \left(2 \omega \left
(t-z\right)\right)\right)
\\
 0 & 0 & 0 & 0
\\
 0 & 0 & 0 & 0
\\
 2   w^{2} \left(1+3 \cos\! \left(2 \omega \left(t-z\right)\right)
\right) & 0 & 0 & -2   w^{2} \left(1+3 \cos\! \left(2 \omega \left
(t-z\right)\right)\right)
\end{array}\right) \label{eq:eqh2}
\end{equation}

The solution to the inhomogeneous field equations consists of the sum of a general solution to the homogeneous equation and a particular solution that depends solely on the first-order perturbation \( h_{\alpha \beta} \). In this analysis, we assume that the second-order perturbation \( \bar{h}^{(2)}_{\alpha \beta} \) also propagates along the positive \( z \)-axis. As a result, the metric components are independent of the transverse coordinates \( x \) and \( y \).
Under this assumption, the solution for the component \( \bar{h}^{(2)}_{00} \) takes the following form:
\begin{equation}
    \bar{h}^{(2)}_{ 0 0 }= F_{0 0 }
\left(-t+z\right)+\frac{  w t \left(2 \omega t+3 \sin\!
\left(2 \omega \left(t-z\right)\right)\right)}{2}
\end{equation} 
the Lorentz gauge condition imposes relations between the components $\bar{h}^{(2)}_{ 0 3}$ and $\bar{h}^{(2)}_{33}$. In particular, we have:
\begin{equation}
    \begin{split}
    &-\frac{\partial}{\partial t} h^{(2)}_{ 0  0} +\frac{\partial}{\partial
    z}h^{(2)}_{ 3  0}\! =0, \\
     &-\frac{\partial}{\partial t} h^{(2)}_{ 0  3}
    +\frac{\partial}{\partial z}h^{(2)}_{ 3  3}\! =0 \label{eq:lorh2}
        \end{split}
\end{equation}
Solving this system leads to the following solutions:
\begin{equation}  
\begin{split}
    &\bar{h}^{(2)}_{0 3 }=2   t \,\omega^{2} z-\frac{3   \sin\!
    \left(2 \omega
    \left(t-z\right)\right) t \omega}{2}+ \frac{3   \cos\! \left(2
    \omega
    \left(t-z\right)\right)}{4}- F_{0 0 }
\left(-t+z\right) +f_{03}\! \left(t\right) \,, \\ 
 &\bar{h}^{(2)}_{3 3 }=F_{0 0 }
\left(-t+z\right) 
    +  \omega^{2} z^{2}-\frac{3   \cos\! \left(2 \omega \left
    (t-z\right)
    \right)}{2}+ \frac{3   \sin\! \left(2 \omega \left(t-z\right)
    \right)
    t \omega}{2} + z \frac{\partial}{\partial t}f_{03}\! \left(t\right)
    +f_{33}\! \left(t\right)\,.
\end{split}
\end{equation} 
Here, the arbitrary function \( \mathit{F}_{00} \) represents the homogeneous solution of the wave equation, while the functions \( f_{03} \) and \( f_{33} \) arise from the gauge constraints~\eqref{eq:lorh2}. All other components of \( \bar{h}^{(2)}_{\alpha\beta} \) are given by homogeneous solutions to the d'Alembert equation.
As in the linear case, the Lorentz gauge condition at second order remains preserved under further coordinate transformations of the form~\eqref{eq:ipt2}, provided the vector fields \( \xi_\mu \) and \( \xi^{(2)}_\mu \) satisfy the wave equations \( \Box \xi_\mu = 0 \) and \( \Box \xi^{(2)}_\mu = 0 \), respectively. The former allows one to impose the TT gauge for \( h_{\alpha\beta} \), while the latter permits a similar gauge fixing at second order.
Thus, we can choose the functions \( \xi^{(2)}_\mu \) in such a way that they do not affect the inhomogeneous part of the solution, while simultaneously ensuring that the homogeneous part of the solution to equation~\eqref{eq:eqh2} also satisfies the transverse-traceless (TT) gauge conditions.
As a result, the homogeneous part can be chosen to be a periodic function, i.e.,
\begin{equation}
    \begin{split}
    &\bar{h}_{(2) 11}=-\bar{h}_{(2) 22} =  \cos( \omega_{2} (t-z) )
    \, ,
    \\
     &\bar{h}_{(2) 12} = \bar{h}_{(2) 21} =  \cos( \omega_{2} (t-z) ) \,.
    \end{split}
\end{equation} 
We introduce a second frequency \( \omega_2 \) in order to distinguish between effects arising from the nonlinear interaction of first-order gravitational waves and those associated with genuinely second-order wave contributions.
To satisfy the gauge constraints and field equations~\eqref{eq:eqh2}, we choose the arbitrary functions \( f_{03} \) and \( f_{33} \) as follows:
\begin{equation}
    \begin{split}
        f_{03} &= -\omega^2 t^2 - \frac{3}{4}\,, \\
        f_{33} &= 2 \omega^2 t^2 + \frac{3}{2}\,.
    \end{split}
\end{equation}
Substituting back into the definition \( h^{(2)}_{\alpha \beta} = \bar{h}^{(2)}_{\alpha \beta} - \frac{1}{2} \eta_{\alpha \beta} \bar{h}^{(2)} \), we obtain the full second-order contribution to the metric tensor. The complete metric \( g_{\mu \nu} \), including both first- and second-order gravitational wave perturbations as given in \eqref{eq:mett2}, takes the form:
\begin{equation}
    \begin{split} \label{eq:exmetric}
        &g_{0 0} = -1+\frac{H^{2} \left(6 t^{2} \omega^{2}-4 t \,\omega^{2} z +2 \omega^{2} z^{2}+6 \sin \! \left(2 \omega  \left(t-z \right)\right) t \omega -3 \cos \! \left(2 \omega  \left(t-z \right)\right)+3\right)}{8}\,, \\
        &g_{0 3} = \frac{H^{2} \left(-3-4 t^{2} \omega^{2}+8 t \,\omega^{2} z -6 \sin \! \left(2 \omega  \left(t-z \right)\right) t \omega +3 \cos \! \left(2 \omega  \left(t-z \right)\right)\right)}{8}\,, \\
        &g_{1 1} = 1  + \frac{H^{2} \cos \! \left(\omega_{2}  \left(t-z \right)\right)}{2}+H \cos \! \left(\omega  \left(t-z \right)\right)-\frac{3 \left(\sin^{2}\left(\omega  \left(t-z \right)\right)\right) H^{2}}{4} -\frac{\left(t-z \right)^{2} \omega^{2} H^{2}}{4}\,, \\
        &g_{2 2} = 1 -\frac{H^{2} \cos \! \left(\omega_{2}  \left(t-z \right)\right)}{2}-H \cos \! \left(\omega  \left(t-z \right)\right)-\frac{3 \left(\sin^{2}\left(\omega  \left(t-z \right)\right)\right) H^{2}}{4}-\frac{\left(t-z \right)^{2} \omega^{2} H^{2}}{4}\,, \\
        &g_{1 2} = \frac{H^{2} \cos \! \left(\omega_{2}  \left(t-z \right)\right)}{2}+H \cos \! \left(\omega  \left(t-z \right)\right)\,,\\
        &g_{3 3} = 1+\frac{H^{2} \left(6 t^{2} \omega^{2}-4 t \,\omega^{2} z +2 \omega^{2} z^{2}+6 \sin \! \left(2 \omega  \left(t-z \right)\right) t \omega -3 \cos \! \left(2 \omega  \left(t-z \right)\right)+3\right)}{8}\,.
    \end{split}
\end{equation} 
It is important to emphasize that the resulting metric tensor does not represent the global structure of spacetime, but is valid only within a region of one wavelength. In other words, the coordinates \( t \) and \( z \) are restricted to the domain \( t \leq 2\pi/\omega \), \( z \leq 2\pi/\omega \).
Furthermore, in this approximation we neglect all terms of order \( H^3 \) and higher. Within this regime, the non-vanishing independent components of the Riemann tensor are presented in the Appendix. The Ricci tensor and scalar curvature vanish:
\begin{equation}
R_{\mu\nu} = 0\,,
\end{equation}
confirming that the spacetime remains a vacuum solution and thus describes free gravitational waves. Moreover, the Kretschmann scalar is also found to vanish.

\section{Second-Order Effects}

\subsection{Geodesic Congruences}

We now consider a congruence of timelike geodesics characterized by a tangent vector field \( U^\alpha \), which satisfies the following conditions:
\begin{equation}
    U_\alpha U^\alpha = -1, \quad U^\alpha_{\ ;\beta} U^\beta = 0,
\end{equation}
where the first relation ensures proper normalization of the four-velocity, and the second indicates that the particles follow geodesics. The nonzero components of the vector \( U^\alpha \) in this spacetime are given by:
\begin{equation} 
\begin{split}
    &U^{0} = 1-\frac{\left(-6 \sin \! \left(2 \omega  \left(t-z \right)\right) t \omega -6 t^{2} \omega^{2}+4 t \,\omega^{2} z -2 \omega^{2} z^{2}+3 \cos \! \left(2 \omega  \left(t-z \right)\right)-3\right) H^{2}}{16} \,, \\ 
    &U^{3} = \frac{3 H^{2} \left(-\cos \! \left(2 \omega  \left(t-z \right)\right)+\omega  \left(\sin \! \left(2 \omega  \left(t-z \right)\right)+\omega  \left(t -2 z \right)\right) t \right)}{8} \,.\\
\end{split}
\end{equation}  
It is now possible to define the transverse metric associated with the geodesic congruence as
\begin{equation}
\tilde{h}_{\mu \nu} = g_{\mu \nu} + U_\mu U_\nu,
\end{equation}
which projects onto the hypersurface orthogonal to the four-velocity \( U^\mu \). The tensor field
\begin{equation}
B_{\alpha \beta} = U_{\alpha ; \beta},
\end{equation}
describes the rate of change of separation between neighboring geodesics. It can be decomposed as
\begin{equation}
B_{\alpha \beta} = \frac{1}{3} \theta \tilde{h}_{\alpha \beta} + \sigma_{\alpha \beta} + \omega_{\alpha \beta},
\end{equation}
where \( \theta \) is the expansion scalar, \( \sigma_{\alpha \beta} \) is the shear tensor, and \( \omega_{\alpha \beta} \) is the rotation tensor.
The expansion scalar is given by the expression
\begin{equation}
\theta = \frac{\omega \left( -2\omega t + \omega z + \sin\left(2\omega(t - z)\right) \right) H^2}{4},
\end{equation}
which arises as a second-order effect and reflects the effective energy content of the gravitational wave. For representative parameters such as \( H = 10^{-20} \) and \( \omega = 100~\text{Hz} \), the magnitude of \( \theta \) is of order \( 10^{-38} \), demonstrating the extreme weakness of the nonlinear tidal expansion induced by the wave.
The nonzero components of the shear tensor are given by:
\begin{equation}
    \begin{split}  \label{eq:shear}
       &\sigma_{xx} =  -\left(\frac{11 \omega  \sin \! \left(2 \omega  \left(t-z \right)\right)}{24} - \frac{\omega_2  \sin \! \left(\omega_2  \left(t-z \right)\right)}{4}+\frac{\left(t -2 z \right) \omega^{2}}{12}\right) H^{2}+\frac{\omega  \sin \! \left(\omega  \left(t-z \right)\right) H}{2}\,, \\ 
        &\sigma_{yy} =  -\left(\frac{11 \omega  \sin \! \left(2 \omega  \left(t-z \right)\right)}{24}+\frac{\omega_2  \sin \! \left(\omega_2  \left(t-z \right)\right)}{4}+\frac{\left(t -2 z \right) \omega^{2}}{12}\right) H^{2}-\frac{\omega  \sin \! \left(\omega  \left(t-z \right)\right) H}{2}\,, \\
       &\sigma_{zz} =  \frac{H^{2} \omega  \left(2 \omega  t -4 \omega  z -\sin \! \left(2 \omega  \left(t-z \right)\right)\right)}{12}\,, \\
       &\sigma_{xy} =\,  \sigma_{yx} =  -\frac{H \left(H \sin \! \left(\omega_{2}  \left(t-z \right)\right) \omega_{2}  +2 \omega  \sin \! \left(\omega  \left(t-z \right)\right)\right)}{4} \,. 
    \end{split}
\end{equation} 
Here, we observe that the components \( \sigma_{xx} \), \( \sigma_{yy} \), and \( \sigma_{zz} \) contain additional terms proportional to \( H^2 \), which originate from the nonlinear contribution of gravitational wave energy as well as second-order waves with frequency \( \omega_2 \). It is also found that the rotation tensor vanishes, i.e., \( \omega_{\alpha \beta} = 0 \), consistent with a hypersurface-orthogonal geodesic flow.
Under these conditions, the Raychaudhuri equation reduces to
\begin{equation}
    \frac{d\theta}{d\tau} = -\omega^2 H^2 \sin^2\left(\omega(t - z)\right),
\end{equation}
which shows that the expansion scalar always decreases or remains constant, confirming the presence of periodic geodesic focusing induced by the wave.

In the context of a null geodesic congruence, we introduce two vector fields \( N^\alpha \) and \( k^\alpha \), satisfying the following normalization conditions:
\begin{equation}
    N_\alpha N^\alpha = 0, \quad k_\alpha k^\alpha = 0, \quad N_\alpha k^\alpha = -1 \,.
\end{equation}
Their components are given by:
\begin{equation}
\begin{split}
    &k^{\mu} = \left( 1+\frac{ H^2 (t+z)^2 \omega^2}{8}, 0, 0, 1-\frac{ H^2 (t+z)^2 \omega^2}{8} \right) \,, \\
    &N_{0} = -\frac{1}{2}+\frac{H^{2} \left(-3 \cos \! \left(2 \omega  \left(t-z \right)\right)+6 \sin \! \left(2 \omega  \left(t-z \right)\right) t \omega +3+\left(5 t^{2}-6 t z +z^{2}\right) \omega^{2}\right)}{16} \,, \\
    &N_{1} = 0\,, N_{2} = 0\,, \\
    &N_{3} = -\frac{1}{2}-\frac{H^{2} \left(-3 \cos \! \left(2 \omega  \left(t-z \right)\right)+6 \sin \! \left(2 \omega  \left(t-z \right)\right) t \omega +3+\left(5 t^{2}-6 t z +z^{2}\right) \omega^{2}\right)}{16}\,.
\end{split}
\end{equation} 
The transverse metric \( \Tilde{H}_{\mu\nu} \) and the tensor field \( B_{\alpha\beta} \) associated with the null geodesic congruence generated by the vector field \( k^\mu \) are defined as
\begin{equation}
    \Tilde{H}_{\mu \nu} = g_{\mu \nu} + k_{\mu} N_{\nu} + k_{\nu} N_{\mu}\,, \quad B_{\alpha \beta} = k_{\alpha ; \beta}
\end{equation}
The evolution tensor \( \Tilde{B}_{\alpha \beta} \), which characterizes the evolution of the cross-sectional area of the congruence, is given by
\begin{equation}
    \Tilde{B}_{\alpha \beta} = \Tilde{H}^{\mu}_{\alpha} \Tilde{H}^{\nu}_{\beta} B_{\mu \nu}\,, \quad  \Tilde{B}_{\alpha \beta} = \frac{1}{2} \Tilde{\theta} \Tilde{H}_{\alpha \beta} + \Tilde{\sigma}_{\alpha \beta} + \Tilde{\omega}_{\alpha \beta}
\end{equation} 
where \( \Tilde{\theta} \) is the expansion scalar, \( \Tilde{\sigma}_{\alpha \beta} \) is the shear tensor, and \( \Tilde{\omega}_{\alpha \beta} \) is the rotation tensor. In the present case, to order \( \mathcal{O}(H^2) \), all these quantities vanish.

\subsection{The Equation of Geodesic Deviation}

The geodesic deviation equation takes the standard form:
\begin{equation}
    \frac{D^2 \xi^{\alpha}}{d \tau^2} = R^{\alpha}_{\ \beta \mu \nu} u^{\beta} u^{\mu} \xi^{\nu}, \qquad \xi^{\alpha}_{\ ;\beta} u^{\beta} = u^{\alpha}_{\ ;\beta} \xi^{\beta},
\end{equation}
where \( \xi^{\alpha} \) is the deviation vector between neighboring geodesics, and \( u^{\alpha} \) is the four-velocity of the reference geodesic.
It is important to note, however, that \( \xi^\mu \) represents a coordinate separation and not the physical (proper) distance between particles. To correctly account for the proper distance, one must evaluate the Riemann tensor in a local inertial frame, also known as the proper detector frame.
While in the linearized theory the Riemann tensor is invariant under infinitesimal gauge transformations, this invariance does not generally hold beyond first order. Therefore, it becomes necessary to construct a local orthonormal tetrad \( e_a^{\ \mu} \) satisfying the conditions:
\begin{equation}
    g_{\mu \nu} e_a^{\mu} e_b^{\nu} = \eta_{ab}, \qquad e^{\mu}_{ a ;\nu} \, u^{\nu} = 0,
\end{equation}
where \( \eta_{ab} = \mathrm{diag}(-1, 1, 1, 1) \) is the Minkowski metric. The second condition ensures that the tetrad is Fermi–Walker transported along the worldline defined by the four-velocity \( u^{\mu} \).
Thus, the nonzero components of the orthonormal tetrad \( e^\mu_a \) can be determined explicitly.
\begin{equation}
\begin{split}
    e_{0}^{\mu} = &  \, U^{\mu} \,,\\ 
    e_{1}^{\mu} = & \Biggl( 0, 1+ \frac{\left(3 H \left(\cos^{2}\left(\omega  \left(t-z \right)\right)\right)-4 \cos \! \left(\omega  \left(t-z \right)\right)+H \left(-2 \cos \! \left(\omega_{2}  \left(t-z \right)\right)+3+\left(t-z \right)^{2} \omega^{2}\right)\right) H}{8} \\
     & -\frac{H  \left(H \cos \! \left(\omega_{2}  \left(t-z \right)\right)+2 \cos \! \left(\omega  \left(t-z \right)\right)\right)}{4},0  \Biggr)\,, \\
     e_{2}^{\mu} = & \Biggl( 0,-\frac{H \left(H \cos \! \left(\omega_{2}  \left(t-z \right)\right)+2 \cos \! \left(\omega  \left(t-z \right)\right)\right)}{4} , \\
     & 1+\frac{H \left(3 H \left(\cos^{2}\left(\omega  \left(t-z \right)\right)\right)+4 \cos \! \left(\omega  \left(t-z \right)\right)+H \left(2 \cos \! \left(\omega_{2}  \left(t-z \right)\right)+3+\left(t-z \right)^{2} \omega^{2}\right)\right)}{8},0  \Biggr) \,, \\
     e_{3}^{\mu} = & \Biggl( -\frac{\left(3 \sin \! \left(2 \omega  \left(t-z \right)\right) t \omega +3+t \left(t -2 z \right) \omega^{2}\right) H^{2}}{8}, 0, 0, \\
     & 1 -\frac{H^{2} \left(6 t^{2} \omega^{2}-4 t \,\omega^{2} z +2 \omega^{2} z^{2}+6 \sin \! \left(2 \omega  \left(t-z \right)\right) t \omega -3 \cos \! \left(2 \omega  \left(t-z \right)\right)+3\right)}{16} \Biggr) \,.
\end{split}
\end{equation}  
Subsequently, the Riemann tensor is projected onto the tetrad frame, yielding physically meaningful components:
\begin{equation}
    \Tilde{R}^{a}_{\ bcd} = R^{\alpha}_{\ \beta \mu \nu} \, e^{a}_{\ \alpha} e^{\beta}_{\ b} e^{\mu}_{\ c} e^{\nu}_{\ d},
\end{equation}
where indices \( a, b, c, d \in \{0, 1, 2, 3\} \) refer to the local orthonormal frame. In particular, the geodesic deviation equation in this frame reduces to:
\begin{equation}
    \frac{d^2 \xi^{a}}{d t^2} = \Tilde{R}^{a}_{\ 0 0 d} \, \xi^{d},
\end{equation}
where \( \xi^a \) denotes the components of the deviation vector in the tetrad basis. The nonvanishing equations take the form:
\begin{equation}
\begin{split}
    \frac{d^2 \xi^{1}}{d t^2} &= - \frac{H}{4} \left( H \omega_2^{2} \cos\left(\omega_2 (t - z)\right) + 2 \omega^{2} \cos\left(\omega(t - z)\right) \right) (\xi^{1} + \xi^{2})\,, \\
    \frac{d^2 \xi^{2}}{d t^2} &= - \frac{H}{4} \left( H \omega_2^{2} \cos\left(\omega_2 (t - z)\right) + 2 \omega^{2} \cos\left(\omega(t - z)\right) \right) (\xi^{1} - \xi^{2})\,.
\end{split}
\end{equation}
These expressions differ from their first-order counterparts by the presence of an additional term proportional to \( H^2 \omega_2^2 \), reflecting a second-order nonlinear correction. Notably, the geodesic deviation remains unaffected by the effective energy of the wave (which scales as \( \omega^2 \)), as it does not enter independently into the equation of deviation.
Assuming \( \omega_2 = \omega \), the amplitude of deviation in a ring of test particles becomes slightly enhanced compared to the linearized theory, with the correction scaling as \( H^2 \). However, for gravitational waves incident on Earth, this correction is many orders of magnitude below current experimental sensitivity and thus remains unobservable.
It is also important to emphasize that, although the shear tensor \eqref{eq:shear} includes a longitudinal component \( \sigma_{zz} \) and compensating modifications in \( \sigma_{xx} \) and \( \sigma_{yy} \) to satisfy the tracelessness condition, the components of the Riemann tensor projected onto the orthonormal frame reveal that the observable expansion and contraction occur exclusively in the transverse \( x \)- and \( y \)-directions.

\subsection{Geodesic Equations in the \( 3+1 \) Formalism}

In the case of linearized gravitational waves, it is well known that a freely falling test body initially at rest remains at rest in the chosen coordinate system after the wave passes. However, when nonlinear effects are taken into account, this behavior can change.
To investigate the coordinate evolution \( x(t), y(t), z(t) \) of a test particle, we adopt the \( 3+1 \) formalism of general relativity~\cite{bib:g31}. In this approach, the spacetime metric~\eqref{eq:exmetric} is decomposed as follows:
\begin{equation}
    g_{\mu \nu} dx^{\mu} dx^{\nu} = -N^2 dt^2 + \gamma_{ij} \left( dx^i + \beta^i dt \right)\left( dx^j + \beta^j dt \right),
\end{equation}
where:
\( N \) is the lapse function, which relates proper time to coordinate time,
\( \beta^i \) is the shift vector, 
\( \gamma_{ij} \) is the induced spatial metric on a spacelike hypersurface of constant \( t \).
In our case, the lapse function and shift vector take the following forms
\begin{equation}
    \begin{split}
        &N= 1-\frac{\left(6 \sin \! \left(2 \omega  \left(t-z \right)\right) t \omega +6 t^{2} \omega^{2}-4 t \,\omega^{2} z +2 \omega^{2} z^{2}-3 \cos \! \left(2 \omega  \left(t-z \right)\right)+3\right) H^{2}}{16}\, ,   \\
         &\beta^{1} = 0 \,, \beta^{2} = 0 \,,\\
         &\beta^3 = -\frac{H^{2} \left(6 \sin \! \left(2 \omega  \left(t-z \right)\right) t \omega -3 \cos \! \left(2 \omega  \left(t-z \right)\right)+3+4 t \left(t -2 z \right) \omega^{2}\right)}{8}.
    \end{split}
\end{equation}
The geodesic equations in the \( 3+1 \) formalism~\cite{bib:geod31} take the following form:
\begin{multline}  \label{eq:g31}
        \ddot{X}^{i}+\frac{1}{N}\left[ K_{jk}\left(
        \dot{X}^{j}+\beta^{j}\right)  \left(
        \dot{X}^{k}+\beta^{k}\right) - 2\left(
        \dot{X}^{j}+\beta^{j}\right)\partial_{j}N-\partial_{t}N
        \right]\dot{X}^{i}+\\+2\left[ D_{j} \beta^{i} - N
        K^{i}_{j}+\frac{\beta^{i}}{N}\left( K_{jk} \beta^{k} -
        \partial_{j}N
        \right)
        \right] \dot{X}^{j} + \left( ^{3}\Gamma^{i}_{jk}+
        \frac{\beta^{i}}{N}K_{jk}
        \right) \dot{X}^{j} \dot{X}^{k} +\\+ N \gamma^{ij} \partial_{j}N-2
        N K^{i}_{j} \beta^{j} \frac{\beta^{i}}{N} \left( K_{jk}
        \beta^{i} \beta^{k} - \partial_{t}N-\beta^{j} \partial_{j} N
        \right) + \partial_{t} \beta^{i} + \beta^{j} D_{j} \beta^{i} =0
\end{multline}
where \( D_i \) denotes the spatial covariant derivative compatible with the three-metric \( \gamma_{ij} \), \( {}^{(3)}\Gamma^{i}_{\ jk} \) are the associated three-dimensional Christoffel symbols, and \( K_{ij} \) is the extrinsic curvature of the spatial hypersurface. The explicit form of \( K_{ij} \), in addition to the system of equations referenced above, is provided in the Appendix.
A numerical integration of the system reveals that, regardless of the amplitude \( H \) and the frequencies \( \omega \) and \( \omega_2 \), a freely falling particle initially at rest does not acquire any displacement in the transverse directions \( x \) and \( y \). This behavior is consistent with the transverse-traceless nature of gravitational waves at leading order.
However, the situation differs along the longitudinal direction \( z \). For example, consider a test particle with initial position and velocity set to zero. Restoring physical units with \( c = 3 \times 10^8~\mathrm{m/s} \), and taking representative values \( H = 10^{-20} \), \( \omega = 100~\mathrm{Hz} \) (the value of \( \omega_2 \) does not affect the motion), the longitudinal displacement over one wave period \( T = 2\pi/\omega \) is approximately:
\[
z \approx 10^{-33}~\mathrm{m}, \qquad \dot{z} \approx 10^{-31}~\mathrm{m/s}.
\]
Although these values are extremely small and currently unobservable, they highlight a nonlinear effect present in the second-order dynamics of gravitational waves.

\newpage
\section{Conclusion}

In this work, using the general framework of tensor expansion via a Taylor series on a differentiable manifold, it was shown that second-order metric perturbations can be made to satisfy the Lorentz gauge, provided an additional vector field \( \xi^{\mu}_{(2)} \) is introduced. This allowed for the construction of differential equations whose solutions correspond to second-order gravitational waves.
Solving these equations yielded a spacetime metric that includes gravitational waves of both first and second order in the wave amplitude \( H \). This metric is valid within one wavelength, and terms of order \( H^3 \) and higher were neglected throughout the analysis.
Within this approximation, congruences of null and timelike geodesics were analyzed. For the null case, the expansion scalar, shear, and rotation tensors vanish identically. For timelike geodesics, the expansion scalar and the Raychaudhuri equation indicate a small contraction effect, of order \( 10^{-38} \), for parameters \( H = 10^{-20} \), \( \omega = 100 \, \text{Hz} \), over a timescale corresponding to a single wave period. The rotation tensor remains zero, while the shear tensor includes additional diagonal components—beyond those expected in linear theory—associated with the gravitational wave’s energy density. 
Upon constructing a parallel-transported orthonormal tetrad and projecting the Riemann tensor onto this frame, it was found that the induced tidal forces act exclusively in the transverse directions, consistent with expectations from linearized theory.

Moreover, by applying the \( 3+1 \) decomposition of spacetime and solving the resulting geodesic equations numerically under zero initial conditions, it was shown that the particle remains stationary in the \( x \) and \( y \) directions, but acquires a small displacement along \( z \). For \( H = 10^{-20} \) and \( \omega = 100 \, \text{Hz} \), this displacement is approximately \( 10^{-32} \, \text{m} \), and the corresponding velocity is \( 10^{-31} \, \text{m/s} \), both negligible from an observational perspective.
However, if the wave amplitude were increased by ten orders of magnitude, the resulting displacement could reach \( 10^{-12} \, \text{m} \), which may become significant. A similar nonlinear displacement effect was reported in~\cite{bib:gribback}, where strong gravitational radiation near the source leads to a net expansion of space. This suggests that due to the inherent nonlinearity of general relativity, such effects may manifest even far from the source though not necessarily extending to asymptotic infinity as memory-like effects.
It is important to emphasize that all results presented here are derived under the assumption of a monochromatic plane gravitational wave with constant amplitude and frequency. In reality, gravitational waves are spherical, with amplitude decreasing with distance and time. Therefore, to obtain more accurate estimates of the nonlinear corrections discussed in this work, one must consider a more general, radiative wave solution.

\textbf{Acknowledgments}  The author gratefully acknowledges partial support from the Theoretical Physics and Mathematics
Advancement Foundation BASIS, grand no.20-1-5-109-1 for financial support.

\section{Appendix} 
Independent components of the Riemann tensor
\begin{equation}
    \begin{split}
    &R_{0103}= R_{0112} = R_{0203} =R_{0212} = R_{0303} = R_{0312} = 
    R_{0313} = R_{0323} = R_{1212} = R_{1213} = R_{1223} = 0 \,, \\
    &R_{0101}=  R_{0113} =  R_{1313} =H^{2} \left(\cos^{2}\left(\omega  \left(t-z \right)\right)\right) \omega^{2}+\frac{H^{2} \cos \! \left(\omega_{2}  \left(t-z \right)\right) \omega_{2}^{2}}{4}+\frac{H \,\omega^{2} \cos \! \left(\omega  \left(t-z \right)\right)}{2} \,,\\
    &R_{0202} =  R_{0223} = R_{2323} =  H^{2} \left(\cos^{2}\left(\omega  \left(t-z \right)\right)\right) \omega^{2}-\frac{H^{2} \cos \! \left(\omega_{2}  \left(t-z \right)\right) \omega_{2}^{2}}{4}-\frac{H \,\omega^{2} \cos \! \left(\omega  \left(t-z \right)\right)}{2} \,, \\
    &R_{0102} =  R_{0123} =  R_{0213} =  R_{1323}= \frac{H \left(H \cos \! \left(\omega_{2}  \left(t-z \right)\right) \omega_{2}^{2}+2 \omega^{2} \cos \! \left(\omega  \left(t-z \right)\right)\right)}{4} \,.
    \end{split} 
\end{equation}
The non-zero components of $K_{ij}$
\begin{equation}
    \begin{split}
    &K_{13}=K_{32}=0 \, , \\
    &K_{12} = \frac{H \left(H \omega_{2}  \sin \! \left(\omega_{2}  \left(t-z \right)\right)+2 \omega  \sin \! \left(\omega  \left(t-z \right)\right)\right)}{4} \, , \\
    &K_{11} = \frac{H \left(\omega  \left(3 H \cos \! \left(\omega  \left(t-z \right)\right)+2\right) \sin \! \left(\omega  \left(t-z \right)\right)+H \left(\omega_{2}  \sin \! \left(\omega_{2}  \left(t-z \right)\right)+\left(t-z \right) \omega^{2}\right)\right)}{4} \,, \\ 
    &K_{22} = \frac{H \left(\omega  \left(3 H \cos \! \left(\omega  \left(t-z \right)\right)-2\right) \sin \! \left(\omega  \left(t-z \right)\right)+\left(-\omega_{2}  \sin \! \left(\omega_{2}  \left(t-z \right)\right)+\left(t-z \right) \omega^{2}\right) H \right)}{4} \,, \\
    &K_{33} = \frac{H^{2} \omega^{2} \left(3 t \cos \! \left(2 \omega  \left(t-z \right)\right)+t +z \right)}{4} \,.
    \end{split}
\end{equation}
The system of equations \eqref{eq:g31}, the dot is responsible for the derivative with respect to
coordinate time $t$
\begin{multline*} \label{eq:sys31}
    \ddot{x} = \Biggl( -\frac{3 \dot{x} \omega  \left(\left(-\frac{{\dot{x}}^{2}}{2}-\frac{{\dot{y}}^{2}}{2}+\frac{11 \dot{z}}{6}-\frac{4}{3}\right) \sin \! \left(2 \omega  \left(z -t \right)\right)+\omega  \cos \! \left(2 \omega  \left(z -t \right)\right) t \left({\dot{z}}^{2}-3 \dot{z}+1\right)\right)}{4}
  + \\  -\frac{\omega_{2}  \left({\dot{x}}^{3}+2 {\dot{x}}^{2} \dot{y}+\left(-{\dot{y}}^{2}+2 \dot{z}-2\right) \dot{x}+2 \dot{y} \left(\dot{z}-1\right)\right) \sin \! \left(\omega_{2}  \left(t-z \right)\right)}{4}- \frac{\omega^{2} \left( \left(t +z \right) \left({\dot{z}}^{2}+1\right)+\left({\dot{x}}^{2}+{\dot{y}}^{2}-\dot{z}\right) \left(t-z \right)\right) \dot{x}}{4} \Biggr) H^2 + \\
   + \frac{\left(-{\dot{x}}^{3}-2 {\dot{x}}^{2} \dot{y}+\left({\dot{y}}^{2}-2 \dot{z}+2\right) \dot{x}-2 \dot{y} \left(\dot{z}-1\right)\right) \omega  \sin \! \left(\omega  \left(t-z \right)\right) H}{2}
  \, ,
\end{multline*}
\begin{multline*}
    \ddot{y} = \Biggl( -\frac{3 \left(\left(-\frac{{\dot{x}}^{2}}{2}-\frac{{\dot{y}}^{2}}{2}+\frac{11 \dot{z}}{6}-\frac{4}{3}\right) \sin \! \left(2 \omega  \left(z -t \right)\right)+\omega  \cos \! \left(2 \omega  \left(z -t \right)\right) t \left({\dot{z}}^{2}-3 \dot{z}+1\right)\right) \omega  \dot{y}}{4}
    + \\  -\frac{\left(-{\dot{y}}^{3}+2 \dot{x} {\dot{y}}^{2}+\left({\dot{x}}^{2}-2 \dot{z}+2\right) \dot{y}+2 \dot{x} \left(\dot{z}-1\right)\right) \omega_{2}  \sin \! \left(\omega_{2}  \left(t-z \right)\right)}{4}-\frac{\dot{y} \left(\left(t +z \right) \left({\dot{z}}^{2}+1\right)+\left({\dot{x}}^{2}+{\dot{y}}^{2}-\dot{z}\right) \left(t-z \right)\right) \omega^{2}}{4}
 \Biggr) H^2 + \\
   + \frac{\left({\dot{y}}^{3}-2 \dot{x} {\dot{y}}^{2}+\left(-{\dot{x}}^{2}+2 \dot{z}-2\right) \dot{y}-2 \dot{x} \left(\dot{z}-1\right)\right) \omega  \sin \! \left(\omega  \left(t-z \right)\right) H}{2}
  \, , 
\end{multline*}
\begin{multline*}
    \ddot{z} = \Biggl(-\frac{3 \left(\left(-2 {\dot{z}}^{2}+\left(\dot{z}-1\right) \left(\frac{{\dot{x}}^{2}}{2}+\frac{{\dot{y}}^{2}}{2}\right)+4 \dot{z}-\frac{3}{2}\right) \sin \! \left(2 \omega  \left(t-z \right)\right)+\omega  t \cos \! \left(2 \omega  \left(t-z \right)\right) \left({\dot{z}}^{3}-4 {\dot{z}}^{2}+5 \dot{z}-1\right)\right) \omega}{4} \\  
     -\frac{\omega_{2}  \left({\dot{x}}^{2}+2 \dot{x} \dot{y}-{\dot{y}}^{2}\right) \left(\dot{z}-1\right) \sin \! \left(\omega_{2}  \left(t-z \right)\right)}{4}- \frac{\left(\left(t +z \right) {\dot{z}}^{3}+\left(t-z \right) \left(\left({\dot{x}}^{2}+{\dot{y}}^{2}+3\right) \left(\dot{z}-1\right)-4 {\dot{z}}^{2}\right)+10 t \dot{z}\right) \omega^{2}}{4}
      \Biggr) H^2 \\ 
      -  \frac{\left(\dot{z}-1\right) \left({\dot{x}}^{2}+2 \dot{x} \dot{y}-{\dot{y}}^{2}\right) \omega \sin \! \left(\omega  \left(t-z \right)\right) H}{2}
  \, .
\end{multline*}


\begin{thebibliography}{150}

\bibitem{bib:gw1}    R. Abbott et al. (LIGO Scientific  Collaboration
and Virgo Collaboration)  \emph{Phys. Rev. Lett.} \textbf{2016},
\mbox{\emph{116}, 061102.}
\bibitem{bib:gw2}   R. Abbott et al. (LIGO Scientific Collaboration, Virgo
Collaboration, and KAGRA Collaboration) \emph{Phys. Rev. D} \textbf{2022},
\mbox{\emph{106}, 062002.}
\bibitem{bib:gw3}  R. Abbott et al. (LIGO Scientific Collaboration,
Virgo Collaboration, and KAGRA Collaboration) \emph{Phys. Rev. D}
\textbf{2022}, \mbox{\emph{106}, 102008.}
\bibitem{bib:gw4} R. Abbott et al. (LIGO Scientific Collaboration, Virgo
Collaboration, and KAGRA Collaboration) Open data from the third
observing run of LIGO, Virgo, KAGRA and GEO
\bibitem{bib:gw5} R. Abbott et al. (LIGO Scientific
Collaboration, Virgo Collaboration, and KAGRA Collaboration) GWTC-3:
Compact Binary Coalescences Observed by LIGO and Virgo During the Second
Part of the Third Observing Run
\bibitem{bib:lisa} Pau Amaro-Seoane et. al. eLISA: Astrophysics and
cosmology in the millihertz regime https://arxiv.org/abs/1201.3621
\bibitem{bib:intgw}  Interaction of gravitational waves with matter A.
Cetoli and C. J. Pethick \emph{Phys. Rev. D} \textbf{2012},
\mbox{\emph{85}, 064036.}
\bibitem{bib:chigw}     Li Fang-Yu et al.  Chinese Phys. B 22 120402
\emph{Chinese Phys. B} \textbf{2013}, \mbox{\emph{22}, 120402.}
\bibitem{bib:romangw}    Romano, J.D., Cornish, N.J. Detection methods
for stochastic gravitational-wave backgrounds: a unified treatment.
Living Rev Relativ 20, 2 (2017).
\bibitem{bib:grishuk}    D Baskaran and L P Grishchuk 2004 Class. Quantum
Grav. 21 4041
\bibitem{bib:dev1}  M. Mohseni,World-line deviation and spinning
particles, Phys. Lett.B 587, 133(2004)
\bibitem{bib:dev2}    M. Heydari-Fard and S. N. Hasani Higher-order
geodesic deviation for charged particles and resonance induced by
gravitational waves International Journal of Modern Physics DVol. 27, No
. 04, 1850042 (2018)
\bibitem{bib:gwingrex1} Bondi H. , Pirani F. A. E. and Robinson I.
1Gravitational waves in general relativity III. Exact plane waves Proc
. R. Soc. Lond. A251519-533,
\bibitem{bib:gwingrex2} Spherical Gravitational Waves Ivor Robinson and
A. Trautman Phys. Rev. Lett. 4, 431
\bibitem{bib:gwingr}   Bondi, H., et al. Gravitational Waves in General
Relativity. VII. Waves from Axi-Symmetric Isolated Systems. Proceedings
of the Royal Society of London. Series A, Mathematical and Physical
Sciences, vol. 269, no. 1336, 1962, pp. 21-52
\bibitem{bib:gwingr2} Sachs, R. Gravitational Waves in General Relativity.
VI. The Outgoing Radiation Condition. Proceedings of the Royal Society
of London. Series A, Mathematical and Physical Sciences, vol. 264, no.
1318, 1961, pp. 309-38
\bibitem{bib:gwingr3} Sachs, R. K. Gravitational Waves in General
Relativity. VIII. Waves in Asymptotically Flat Space-Time. Proceedings
of the Royal Society of London. Series A, Mathematical and Physical
Sciences, vol. 270, no. 1340, 1962, pp. 103-26.
\bibitem{bib:gwingr4} Asymptotic Symmetries in Gravitational Theory R. 
Sachs Phys. Rev. 128, 2851  1962
\bibitem{bib:meggori} M. Maggiore,
Gravitational Waves: Volume1: Theory and Experiments,(Oxford University
Press., NewYork,2008 
\bibitem{bib:non1} Towards a theory of nonlinear gravitational
waves: A systematic approach to nonlinear gravitational perturbations in
the vacuum Andrzej Rostworowski Phys. Rev. D 96, 124026 2017
\bibitem{bib:non2} Gleiser, R., Nicasio, C., Price, R.,  Pullin, J.
(1996). Second-order perturbations of a Schwarzschild black hole.
Classical and Quantum Gravity, 13 (10)   
\bibitem{bib:non3} Second-order quasinormal mode of the
Schwarzschild black hole Hiroyuki Nakano and Kunihito Ioka Phys. Rev. D 
76, 084007 2007
\bibitem{bib:non4} Nicholas Loutrel, Justin L. Ripley, Elena Giorgi,
and Frans Pretorius Phys. Rev. D 103, 104017  2021
\bibitem{bib:g1}  M. Bruni, S. Matarrese, S. Mollerach and S.
Sonego, Perturbations of spacetime: gauge transformations and gauge
invariance at second order and beyond,Classical and Quantum Gravity,
vol. 14, no. 9, pp. 2585-2606, 1997
\bibitem{bib:g2} K. Nakamura, Gauge-invariant formulation of
second-order cosmological perturbations, Physical Review D, vol. 74, no
. 10, Article ID 101301, 2006.
\bibitem{bib:g3} S. Sonego and M. Bruni, Gauge dependence in the
theory of non-linear spacetime perturbations, Communications in
Mathematical Physics vol. 193, no. 1, pp. 209-218, 1998.
\bibitem{bib:g4} Relativistic second-order perturbations of the
Einstein-de Sitter universe Sabino Matarrese, Silvia Mollerach, and
Marco Bruni Phys. Rev. D 58, 043504  1998
\bibitem{bib:g31}  E. Gourgoulhon 3+1 Formalism in General
Relativity Springer-Verlag Berlin Heidelberg 2012
\bibitem{bib:geod31} 3+1 geodesic equation and images in numerical
spacetimes Frederic H. Vincent(Meudon Observ. and LUTH, Meudon), Eric
Gourgoulhon(LUTH, Meudon), Jerome Novak(LUTH, Meudon) Aug, 2012 20 pages
Published in: Class.Quant.Grav. 29 (2012) 245005
\bibitem{bib:gribback}  A. A. Grib and Yu. V. Pavlov Back reaction
of the gravitational radiation on the metric of spacetime  International
Journal of Modern Physics D Vol. 27, No. 07, 1850071 (2018)
\end{thebibliography}
\end{document}